# Models for Planning and Simulation in Computer Assisted Orthognatic Surgery

Matthieu Chabanas[1], Christophe Marecaux[1,2], Yohan Payan[1] and Franck Boutault[2]

[1] TIMC/IMAG laboratory, UMR CNRS 5525, Université Joseph Fourier, Grenoble, France
Institut Albert Bonniot - 38706 La Tronche cedex, France
`Matthieu.Chabanas@imag.fr`
[2] CCRAO Laboratory – Université Paul Sabatier – Toulouse, France
Service de Chirurgie Maxillo-faciale, Hôpital Purpan - 31059 Toulouse cedex, France

**Abstract.** Two aspects required to establish a planning in orthognatic surgery are addressed in this paper. First, a 3D cephalometric analysis, which is clinically essential for the therapeutic decision. Then, an original method to build a biomechanical model of patient face soft tissue, which provides evaluation of the aesthetic outcomes of an intervention. Both points are developed within a clinical application context for computer aided maxillofacial surgery.

## 1 Introduction

Orthognathic surgery attempts to establish normal aesthetic and functional anatomy for patients suffering from dentofacial disharmony. In this way, current surgery aims at normalize patients dental occlusion, temporo mandibular joint function and morphologic appearance by repositioning maxillary and mandibular skeletal osteotomized segments. Soft tissue changes are mainly an adaptation of these bones modifications.

The cranio-maxillofacial surgeon determines an operative planning consisting in quantitative displacement of the skeletal segments (maxillar and/or mandibular) for estimating normal position. Real clinical problems are to respect temporo mandibular joint functional anatomy and to predict soft tissue changes. This last point is important on one part for surgeon as the final soft tissue facial appearance might modify the operative planning and on the other part for the patient who expects a reliable prediction of his post operative aesthetic appearance.

In current practice, the orthognatic surgical planning involves several multimodal data: standard radiographies for bidimensional cephalometric analysis, plaster dental casts for orthodontic analysis, photographs and clinical evaluation for anthropometric measurements. In comparison with normative data set and according to orthodontic and cephalometric analysis, the surgeon simulate dental arch displacements on plaster casts to build some resin splints as reference of the different occlusion stages. These splints are used during surgery to guide maxillary and mandibular osteotomies repositioning. No reliable per operative measurement guaranties defined planning respect. This difficult and laborious process might be responsible of imprecision and requires a strong experience. Medical imaging and computer assisted surgical technologies may improve current orthognatic protocol as an aid in diagnostic, surgical planning and surgical intervention.

This work presents our experience in computer assisted orthognatic surgery. First, we remind sequences of a computer aided cranio-maxillofacial protocol as defined in literature stressing on missing points for a clinical application of these techniques. Then, we address the two points which are, in our minded, the two main remaining problems in this way: a 3D cephalometric analysis and a postoperative facial soft tissue appearance prediction.

## 2  Computer assisted cranio-maxillofacial surgery

The different steps of a computer-aided protocol in cranio-maxillofacial surgery are well defined in the literature [1,2]. They can be summarised in 7 points:

1. CT data acquisition, with computer generated 3D surface reconstruction
2. Three-dimensional cephalometric analysis and dental occlusion analysis for clinical diagnosis and operative planning
3. Surgical simulation, including osteotomies and real time mobilisation of the bone segments with 6 degrees of freedom
4. Prediction of the facial soft tissue deformation according to the repositioning of the underlying bone structures
5. Validation of the surgical planning according to the soft tissue simulations
6. Data transfer to the operative room and per-operative computer aided navigation, to ensure the established planning is accurately respected
7. Evaluation of the surgical outcomes

Different stages of this computer aided protocol for cranio-maxillofacial surgery have been addressed in the literature.

Three dimensional cephalometric analysis, despite being essential for planning decision, has been studied very little so far. A previous work, proposed by our group [3], was an extension of a 2D cephalometry (from Delaire) used for osteotomized segments repositioning. However, cephalometric and orthodontic planning were made in traditional way (on 2D standard teleradiography and plaster dental casts). The most interesting and original work was proposed by Treil [4]. He introduces a new cephalometry based on CT scanner imaging, anatomic landmarks and mathematic tools (maxillofacial frame and dental axis of inertia) for skeletal and dental analysis. However, in our point of view, this cephalometric analysis is not relevant for operative planning and computer guided surgery.

Most of the existing works deal with the interactive simulation of a surgical procedure on 3D skull computer generated models [5,6]. Physical models were also developed to evaluate the aesthetic outcomes resulting from underlying bone repositioning [6,7,8,9]. However, despite their evident scientific interest, most of these work cannot be used in clinical practice, since the bone simulations are not clinically relevant.

Few works exist about per-operative guiding in cranio-maxillofacial surgery [10,11,12].

According to us, none of the working groups consider the whole computer aided sequence. Our group strives to develop a facial skeletal model for cephalometry and osteotomy simulation, and a finite element model of the facial soft tissues for post operative aesthetic appearance simulation. Moreover, we have already addressed bone segmentation, mobilisation and guidance for orthognatic surgery in previous works [10,3].

## 3   3D cephalometry : a morphometric analysis

A complete computer aided cranio-maxillofacial surgery sequence requires a bone skull model that enables the medical diagnostic, supports the surgical bone osteotomies simulation, integrates the post operative facial soft tissues deformation prediction and can be used as interface in computer guided surgery.

To be accepted by medical community, this model must be coherent from an anatomical, physiological and organ genetic point of view. A 3D cephalometric tool as an aid in diagnostic is admitted as useful [1,3,6]. 3D CT scanner imaging is already currently used to apprehend the difficult three dimensional part of this pathology. However, there is no relevant direct three dimensional analysing method of these images.

A reliable cephalometry requires defining a referential for facial skeleton orientation, used for intra and inter patient measurements reproducibility and for quantification of bone displacement, and a facial morphologic analysis for treatment planning decision in comparison to a norm determined as "equilibrated" face.

This model should be able to be segmented for simulation as in a surgical procedure. The finite element facial soft tissue model described in section 4 should also be integrated.

### 3.1 Referential definition

We propose an invariant, reproducible, orthogonal referential, defined by 3 planes (figure 1). An horizontal plane close from cranio basal planes of previous 2D cephalometries and from the horizontal vestibular plane defined as the craniofacial physiologic plane. Its construction uses anatomic reliable landmarks: head of right and left mallei and the middle point between both supraorbital foramina. The medial sagittal and frontal planes are orthogonal to the horizontal plane, and contain the middle point of both head mallei. As defined, this referential is independent from the analysed operated facial skeleton.

The x, y and z coordinates of each voxel are transferred from the original CT scanner referential to this new referential. These normalised coordinates allow location or measurement comparison between two patients or in the same one across time.

### 3.2 Maxillofacial framework for skull analysis

The cephalometry definition requires both a maxillofacial frame for morphologic analysis and a norm, quantitative or qualitative, defined as an ideal for a pleasant equilibrated face. The operative planning is defined by differences between current patient state and this norm.

We propose a maxillofacial frame (figure 1) composed of 15 anatomic reliable landmarks and 9 surfaces [13]. Mathematical tools allow metric, angular and surfacic measurements. Contrary to traditional 2D cephalometry, these are direct values and not measurements between projected and constructed points on a sagittal radiography.

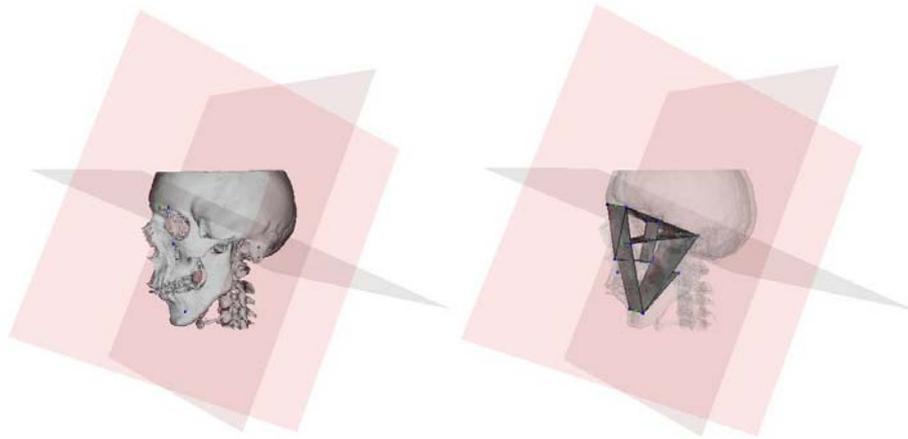

Figure 1: The craniofacial referential, and a 3D analysis example.

## 4 Finite Element model of the face soft tissue

### 4.1 Methodology

Different face models have been developed for simulating maxillofacial surgery outcomes. Although the first ones were based on discrete mass-spring structures [7], most of them use the Finite Element Method to resolve the mechanical equations describing soft tissue behavior [6,8,9]. These models are based on a 3D mesh, generated from patient CT images using automatic meshing methods. Such algorithms are not straightforward in this case, as the boundary of the face soft tissue, i.e. the skin and skull surfaces, must be semi-automatically segmented, which is time-consuming and cannot be used in clinical routine. Moreover, these meshes are composed of tetrahedral elements, less efficient than hexahedral ones in terms of accuracy and convergence.

Our methodology consists, first, in manually building one "generic" model of the face, integrating skin layers and muscles. Then, the mesh of this generic model is conformed to each patient morphology, using an elastic registration method and patient data segmented from CT images. The automatically generated patient mesh has then to be regularized in order to perform Finite Element analysis.

### 4.2 Patient mesh generation

A volumetric mesh was manually designed, representing soft tissue of a "standard" human face [14]. It is composed of two layers of hexahedral elements representing the dermis and hypodermis (figure 2). Elements are organized within the mesh so that the main muscles responsible of facial mimics are clearly identified.

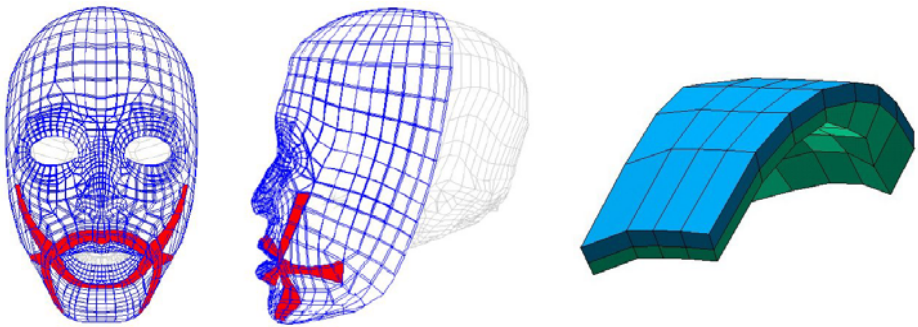

Figure 2. The generic 3D mesh, with embedded main facial muscles

The generic mesh is adapted to each patient morphology using the Mesh-Matching algorithm [15]. This method, based on the Octree Spline elastic registration algorithm [16], computes a non-rigid transformation between two 3D surfaces. The external skin and skull surfaces of the patient are automatically built out of CT images [17]. Then, the patient mesh is generated in two steps (figure 3) :
1. An elastic transformation is computed to fit the *external* nodes of the generic model to the patient *skin* surface, then applied to all the nodes of the mesh.
2. Another transformation is thus calculated between the *internal* nodes of the mesh and the patient *skull* surface. This second transformation is applied to non-fixed internal nodes, i.e. not located in the lips and cheeks area.

A mesh conformed to the specific patient morphology is then available, still integrating the skin and muscles structures.

Since nodes of the mesh are displaced during the registration, some elements can be geometrically distorted. If an element is too distorted, the "shape function" that maps it to the reference element in the Finite Element method cannot be calculated. An automatic algorithm was thus developed to correct these mesh irregularities, by slightly displacing some nodes until every element is regular [18]. Therefore, a regularized patient mesh is obtained, which can be used for Finite Element analysis.

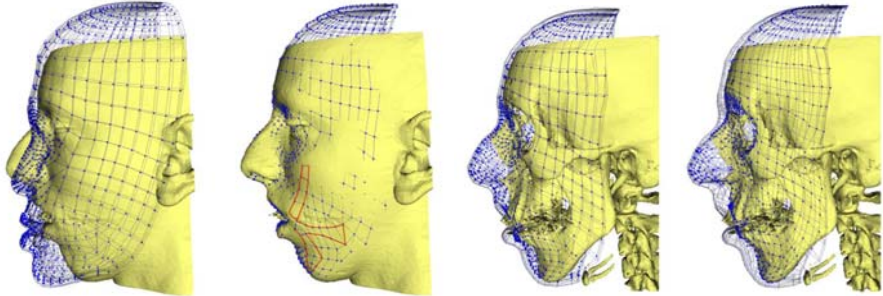

Figure 3. External nodes of the generic mesh are non-rigidly matched to the patient skin surface (left). Then, internal nodes are fitted to the patient skull surface (right). Muscles are still integrated in the new patient mesh

### 4.5 Mechanical properties and boundary conditions

In a first step, simple modeling assumptions are assumed, with linear elasticity and small deformation hypothesis [14]. The anisotropy of the face due to muscular organization is taken into account by setting linear transverse elasticity in the muscles fibers directions. As boundary conditions, internal nodes are rigidly fixed to the skull model, except in the lips and cheeks area. To simulate bone repositioning, nodes fixed to the mandible or maxilla are displaced according to the surgical planning. Muscular activation can also be simulated to produce facial mimics [14].

### 4.6 Results

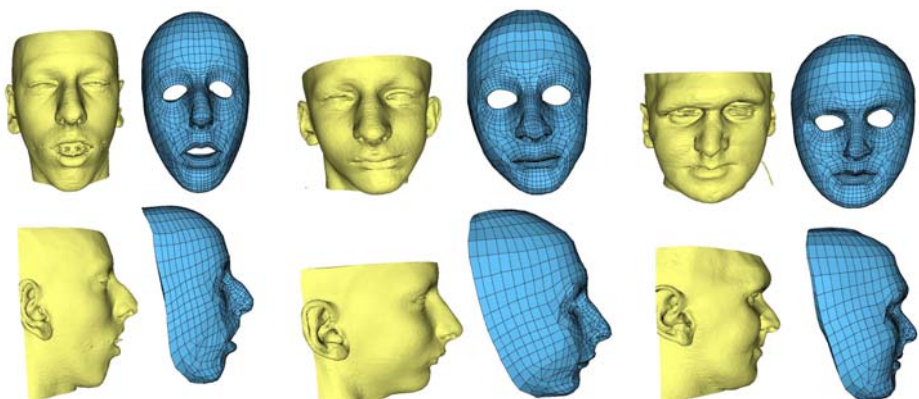

Figure 4. Three models of patients with different morphologies. Each model is quasi-automatically built in about 15 minutes

This mesh generation method was successfully used to build six models of patients with different morphologies. Three of them are presented in figure 4. First results of simulation, carried on the patient presented in figure 3, are shown in figure 5.

The accuracy, given by the matching algorithm, is under 1mm. Although, one of the main advantage of this straightforward and easy to use method is the time required to build a patient model. It is almost automatic, and does not require the interactive definition of landmarks on patient data. The only thing the user has to check is the quality of the marching cubes reconstruction, then the initial position of the generic mesh with respect to the patient skin surface before the registration. The total reconstruction time for a patient model is 15 minutes in mean, principally for the Marching-cube and the mesh regularization computation. Hence, this model generation method is suitable to be routinely used by a surgeon in the elaboration of a surgical planning.

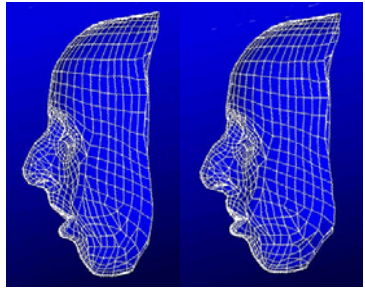

Figure 5. Simulation of soft tissue deformation resulting from mandible and maxilla repositioning

### 4.7 Validation protocol

A primary point when using biomechanical models is their validation, especially in a quantitative way. Since our modeling has been designed in the framework of computer-aided maxillofacial surgery, the simulations of soft tissue deformation will be validated within this framework, using the developed clinical application. Post-operative CT data will be acquired, initially for at least two patients. Three steps will therefore be carried out:
1. The first point is to determine the surgical displacements, in the referential defined in section 3, from anatomical landmarks located on maxilla and mandible.
2. These displacements will then be inputted in the model, to simulate the surgical outcomes in terms of bone repositioning, and therefore soft tissue deformation.
3. Finally, these simulations will be quantitatively compared to the real post-operative appearance of the patient.

Once these quantitative measurements are available, the biomechanical model could be improved (large deformations, non linear law) to enhance the simulations quality.

## 5  Conclusion

New concepts of 3D cephalometry and soft tissue prediction were introduced for computer aided techniques and surgical practice. Current research concerns clinical validation of both models and their integration in a complete protocol.

## References

1. Cutting, C., Bookstein, F.L., Grayson, B., Fellingham, L., Mc Carthy, J.G., 3D computer-assisted design of craniofacial surgical procedures: optimization and interaction with cephalometric and CT-based models. Plast Reconstr Surg, 77(6), pp. 877-885, 1986.
2. Altobelli DE, Kikinis R, Mulliken JB, Cline H, Lorensen W, Jolesz F, Computer-assisted three-dimensional planning in craniofacial surgery. Plast Reconstr Surg 92: 576-587, 1993.
3. Bettega G., Payan Y., Mollard B., Boyer A., Raphaël B. and Lavallée S., A Simulator for Maxillofacial Surgery Integrating 3D Cephalometry and Orthodontia, Journal of Computer Aided Surgery, vol. 5(3), pp. 156-165, 2000.
4. Treil J., Borianne Ph., Casteigt J., Faure J., and Horn A.J., The Human Face as a 3D model: The Future in Orthodontics, World Journal of Orthodontics, vol 2(3), pp. 253-257, 2001.
5. Barré, S., Fernandez, C., Paume, P., Subrenat, G., 2000. Simulating Facial Surgery. Proc. of the IS&T/SPIE Electronic Imaging, vol. 3960, pp. 334-345
6. Zachow, S. ; Gladilin, E. ; Zeilhofer, H.-F. ; Sader, R.: Improved 3D Osteotomy Planning in Cranio-Maxillofacial Surgery. MICCAI'2001, Springer Verlag 2208, pp. 473-481, 2001.
7. Lee Y., Terzopoulos D. and Waters K., Realistic Modeling for Facial Animation, SIGGRAPH'95, pp. 55-62, 1995.
8. Keeve E., Girod S., Kikinis R., Girod B., Deformable Modeling of Facial Tissue for Craniofacial Surgery Simulation, J. Computer Aided Surgery, 3, pp. 228-238, 1998.
9. Schutyser P., Van Cleynenbreugel J., Ferrant M., Schoenaers J., Suetens P., Image-Based 3D Planning of Maxillofacial Distraction Procedures Including Soft Tissue Implications, MICCAI'2000, Springer-Verlag 1935, pp. 999-1007, 2000.
10. Bettega G, Dessenne V, Cinquin P, Raphaël B, Computer assisted mandibular condyle positioning in orthognatic surgery. J. Oral Maxillofac. Surg. 54(5): 553-558, 1996.
11. Marmulla R., Niederdellmann H., Surgical planning of computer-assisted repositioning osteotomies. Plast Reconst Surg; 104:938-944, 1999.
12. Schramm A, Gelldrich NC, Naumann S, Buhner U, Schon R and Schmelzeisen R., Non-invasive referencing in computer assisted surgery. Med Biol Eng Comp, 37:644-645, 1999.
13. Chabanas M., Marecaux Ch., Payan Y. and Boutault F., Computer Aided Planning For Orthonatic Surgery, Computer Assisted Radiology and Surgery, CARS'2002.
14. Chabanas M. and Payan Y., A 3D Finite Element model of the face for simulation in plastic and maxillo-facial surgery, MICCAI'2000, Springer-Verlag 1935, pp.1068-1075, 2000.
15. Couteau B., Payan Y., Lavallée S., The Mesh-Matching algorithm: an automatic 3D mesh generator for finite element structures, J. of Biomechanics, vol. 33/8, pp. 1005-1009, 2000.
16. Szeliski R., Lavallée S., Matching 3-D anatomical surfaces with non-rigid deformations using octree-splines, J. of Computer Vision, 18(2), pp. 171-186, 1996.
17. Lorensen W.E, Cline H.E., Marching Cube: A High Resolution 3D Surface Construction Algorithm. ACM Computer Graphics 21:163-169, 1987.
18. Luboz V., Payan Y., Swider P., Couteau B., Automatic 3D Finite Element Mesh Generation: Data Fitting for an Atlas, Proceedings of the Fifth Int. Symposium on Computer Methods in Biomechanics and Biomedical Engineering, CMBBE'2001.